%% file: main.tex
\documentclass[10pt,journal,compsoc]{IEEEtran}

\usepackage{cite}
\usepackage{amsmath,amssymb,amsfonts}
\usepackage{algorithmic}
\usepackage{graphicx}
\usepackage{textcomp}
\usepackage{xcolor}
\usepackage{tcolorbox}
\usepackage{multirow}
\usepackage{colortbl}
\usepackage{fixmath}
\usepackage{hyperref}
\usepackage{amsmath,amssymb,amsfonts}
\usepackage[ruled,linesnumbered,lined,noend,vlined]{algorithm2e}
\usepackage{algorithmic} 
\usepackage{enumitem}
\usepackage{ragged2e}
\definecolor{light-gray}{gray}{0.95}
\SetKw{Break}{break}
\usepackage[switch]{lineno}

\newcommand{\code}[1]{\colorbox{light-gray}{\texttt{#1}}}
\newcommand{\tool}{\textsf{AnchorCoder}}

\newcommand{\defeq}{\mathrel{\mathop:}=}

\newcommand{\domain}[1]{\mathbb{R}^{#1}}

\def\BibTeX{{\rm B\kern-.05em{\sc i\kern-.025em b}\kern-.08em
		T\kern-.1667em\lower.7ex\hbox{E}\kern-.125emX}}
\begin{document}

\title{Anchor Attention, Small Cache: Code Generation with Large Language Models}

\author{Xiangyu Zhang,
        Yu Zhou,
        Guang Yang,
        Harald C. Gall,
        Taolue Chen
        
\IEEEcompsocitemizethanks{
\IEEEcompsocthanksitem Xiangyu Zhang is with the College of Computer Science and Technology, 
Nanjing University of Aeronautics and Astronautics, Nanjing, China.
E-mail: zhangx1angyu@nuaa.edu.cn
\IEEEcompsocthanksitem Yu Zhou (Corresponding author) is with the College of Computer Science and Technology, 
Nanjing University of Aeronautics and Astronautics, Nanjing, China.
E-mail: zhouyu@nuaa.edu.cn
\IEEEcompsocthanksitem Guang Yang is with the College of Computer Science and Technology, Nanjing University of Aeronautics and Astronautics, 
Nanjing, China.
E-mail: novelyg@outlook.com
\IEEEcompsocthanksitem Harald C. Gall is with the University of Zurich, Zurich, Switzerland. 
E-mail: gall@ifi.uzh.ch
\IEEEcompsocthanksitem Taolue Chen (Corresponding author) is with School of Computing and Mathematical Sciences, Birkbeck, University of London, UK. 
E-mail: t.chen@bbk.ac.uk
}

\thanks{Manuscript received xx, 00,2024; revised xxxx xx, xxxx.}}

\markboth{Preprint.}%
{Shell \MakeLowercase{\textit{et al.}}: Bare Demo of IEEEtran.cls for Computer Society Journals}

\IEEEtitleabstractindextext{
\begin{abstract}
\justifying
The development of large language models (LLMs) has revolutionized automated code generation. However, their high demand of computation resources 
has hindered a broader deployment and raised environmental concerns. 
A common strategy for diminishing computational demands is to cache Key-Value (KV) states from the attention mechanism which is adopted predominately by mainstream LLMs. It can mitigate 
the need of repeated attention computations, but brings significant memory overhead. Current practices in NLP often use sparse attention 
which may, unfortunately, lead to substantial inaccuracies, or hallucinations, in code generation tasks. In this paper, we analyze the attention weights distribution within code generation models via an empirical study, uncovering a sparsity pattern, i.e., the aggregation of information at specific anchor points. Based on this observation, we propose a novel approach, AnchorCoder, which features token-wise anchor attention designed to extract and compress the contextual information, and layer-wise anchor attention enabling cross-layer communication to mitigate the issue of excessive superposition caused by the compression. The extensive experiments across multiple benchmark datasets confirm the effectiveness of AnchorCoder, which can consistently achieve a significant (at least 70\%) reduction in KV cache requirements, while preserving the majority of model's performance. 

\justifying
\end{abstract}

}

\maketitle
\IEEEdisplaynontitleabstractindextext
\IEEEpeerreviewmaketitle


\input{sections/1.introduction.tex}
\input{sections/3.empirical.tex}
\input{sections/4.method.tex}

\input{sections/5.setup.tex}
\input{sections/6.result.tex}

\input{sections/7.threat.tex}
\input{sections/2.background.tex}
\input{sections/8.conclusion.tex}


\section*{Acknowledgment}
This work is supported by the National Natural Science Foundation of China (No.\ 62372232), the Fundamental Research Funds for the Central Universities (No. NG2023005), and the Collaborative Innovation Center of Novel Software Technology and Industrialization. T. Chen is partially supported by an overseas grant from the State Key Laboratory of Novel Software Technology, Nanjing University (KFKT2023A04). 

\bibliographystyle{IEEEtran}
\bibliography{acmart}

\begin{IEEEbiography}[{\includegraphics[width=1in,height=1.25in,clip]{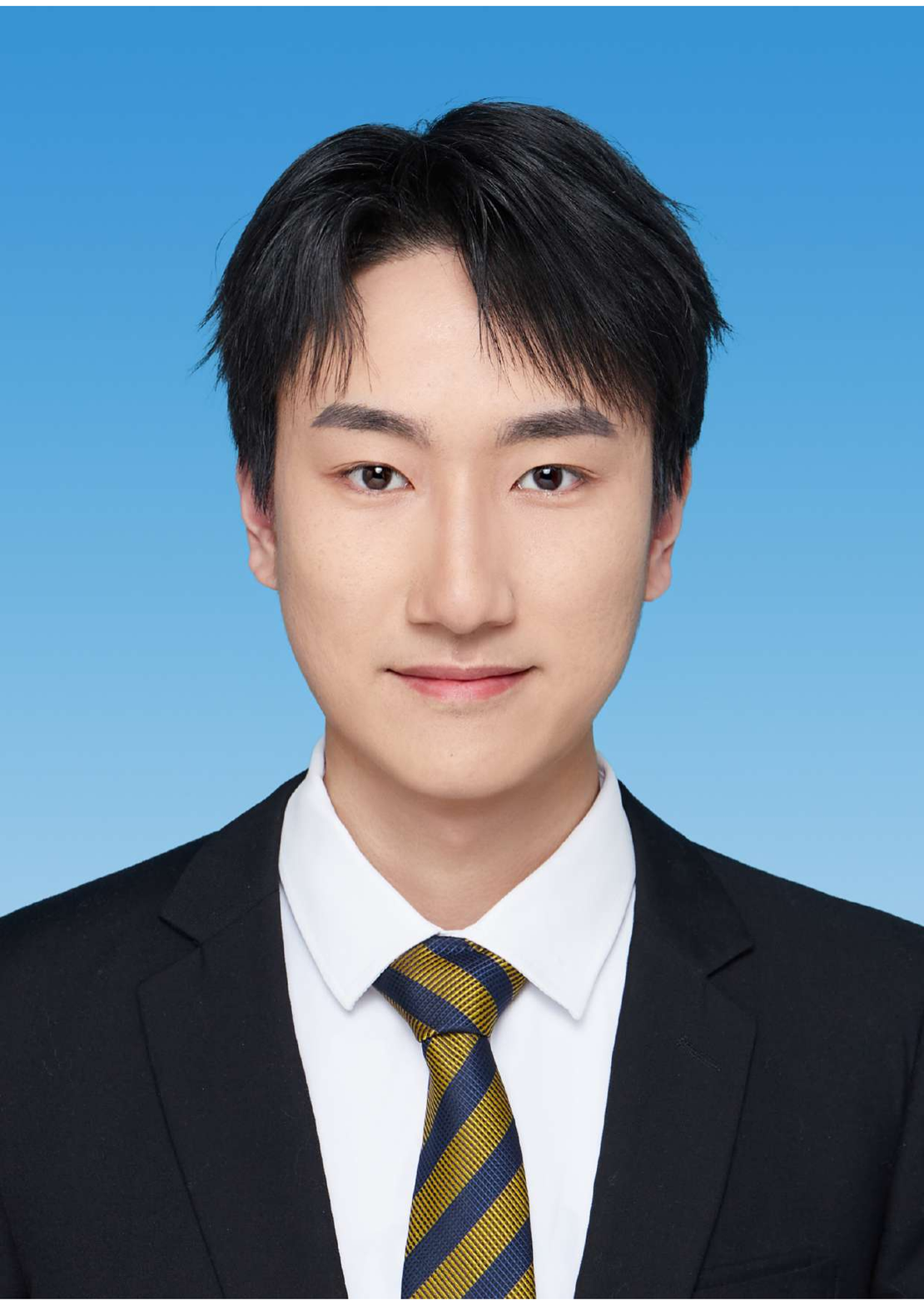}}]{Xiangyu Zhang}
is currently pursuing a Master's degree at the College of Computer Science and Technology of Nanjing University of Aeronautics and Astronautics. His research interests include code generation and model interpretability.
\end{IEEEbiography}

\begin{IEEEbiography}[{\includegraphics[width=1in,height=1.25in,clip]{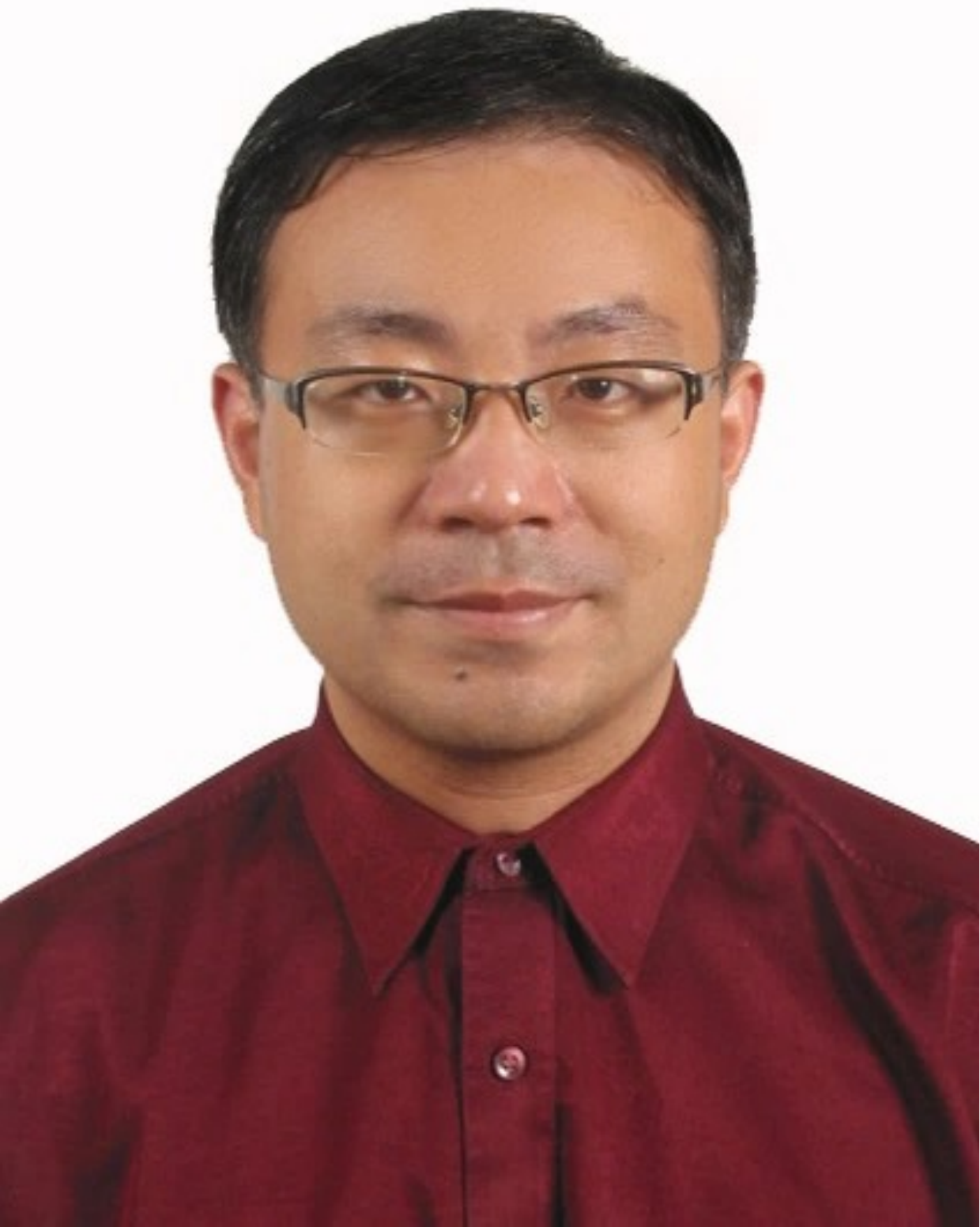}}]{Yu Zhou}
is a full professor in the College of Computer Science and Technology at Nanjing University of Aeronautics and Astronautics (NUAA). He received his BSc degree in 2004 and PhD degree in 2009, both in Computer Science from Nanjing University China. Before joining NUAA in 2011, he conducted PostDoc research on software engineering at Politechnico di Milano, Italy. From 2015-2016, he visited the SEAL lab at University of Zurich Switzerland, where he is also an adjunct researcher. His current research interests mainly generative models for software engineering, software evolution analysis, mining software repositories, and reliability analysis. He has been supported by several national research programs in China. More information about him can be found at: \url{https://csyuzhou.github.io/}.

\end{IEEEbiography}

\begin{IEEEbiography}[{\includegraphics[width=1in,height=1.25in,clip]{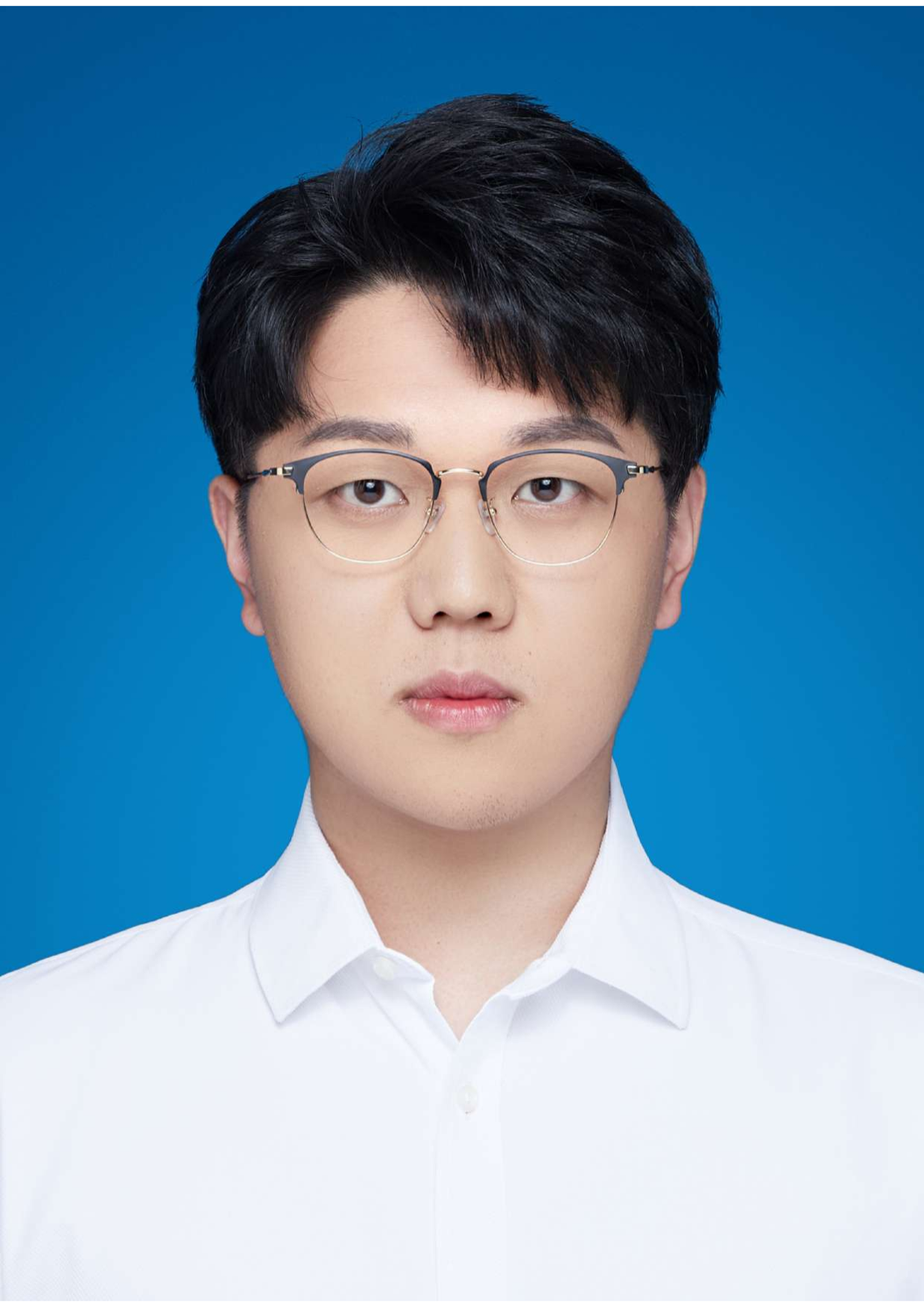}}]{Guang Yang}
received the M.D. degree in computer technology from Nantong University, Nantong, in 2022. Then he is currently pursuing the Ph.D degree at Nanjing University of Aeronautics and Astronautics, Nanjing.
His research interest is AI4SE and he has authored or coauthored more than 20 papers in refereed journals or conferences, such as ACM Transactions on Software Engineering and Methodology (TOSEM), Empirical Software Engineering, Journal of Systems and Software, International Conference on Software Maintenance and Evolution (ICSME), and International Conference on Software Analysis, Evolution and Reengineering (SANER).
More information about him can be found at: \url{https://ntdxyg.github.io/}
\end{IEEEbiography}
\begin{IEEEbiography}[{\includegraphics[width=1in,clip,keepaspectratio]{figures/Author/gall.pdf}}]{Harald C. Gall}  is Dean of the Faculty of Business, Economics, and Informatics at the University of Zurich. He is professor of software engineering in the Department of Informatics. He held visiting positions at Microsoft Research in Redmond, USA, and University of Washington in Seattle, USA. His research interests are software evolution, software architecture, software quality, and cloud-based software engineering. Since 1997, he has worked on devising ways in which mining repositories can help to better understand and improve software development.
\end{IEEEbiography}
\begin{IEEEbiography}[{\includegraphics[width=1in,clip,keepaspectratio]{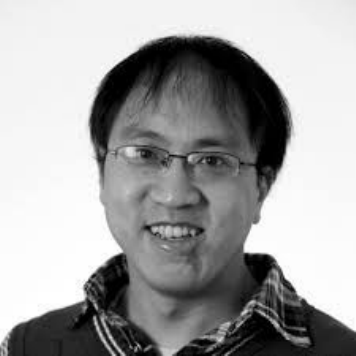}}]{Taolue Chen} received the Bachelor and Master degrees from Nanjing University, China, both in computer science. He was a junior researcher (OiO) at the Centrum Wiskunde \& Informatica (CWI) and acquired the PhD degree from the Vrije Universiteit Amsterdam, The Netherlands. He is currently a senior  lecturer at the School of Computing and Mathematical Sciences, Birkbeck, University of London. He had been a postdoctoral researcher at University of Oxford (UK)  and University of Twente (NL). His research areas include Software Engineering, Programming  Language and Verification. His present research focus is on the border of software engineering and machine learning. 
He has published about 150 papers in journals and conferences such as POPL, LICS, CAV, OOPSLA, ICSE, ESEC/FSE, ASE, ETAPS (TACAS, FoSSaCS, ESOP, FASE), NeurIPS, ICLR, IJCAI, AAAI, EMNLP and IEEE Transactions on Software Engineering (TSE), ACM Transactions on Software Engineering and Methodology (TOSEM), Empirical Software Engineering, ACM Transactions on Computational Logic (TOCL), Information and Computation, Logical Methods in Computer Science. He won the Best Paper Award of SETTA’20, the 1st Prize in the CCF Software Prototype Competition 2022, the QF\_Strings (Single Query Track) at the International Satisfiability Modulo Theories Competition 2023, and ACM SIGSOFT Distinguished Paper Award. He has served  editorial board or program committee  for various international journals and conferences. More information about him can be found at \url{https://chentaolue.github.io/}.
\end{IEEEbiography}

\end{document}

%% file: sections/1.introduction.tex
\section{Introduction} \label{sect:intro}
Automated generation of code that aligns with user intentions poses a significant and enduring challenge in software engineering~\cite{chen2021evaluating, li2022competition, fried2022incoder}. In recent years, the tremendous progress in deep learning and NLP, especially the advent of Large Language Models (LLMs~\cite{huang2023chatgpt,yao2024survey}), has revolutionized the research of automated code generation
~\cite{OpenAI2023GPT4TR,hou2023large}. LLMs for code, e.g., CodeGen~\cite{nijkamp2022codegen}, CodeLlama~\cite{roziere2023code} and CodeGeeX~\cite{zheng2023codegeex}, have showcased impressive proficiency in writing code, boosting the productivity of developers across various programming environments remarkably~\cite{xu2022ide}.

Almost all mainstream LLMs (including those for code which are the main focus of the current paper) adopt the Transformer architecture~\cite{vaswani2017attention}, which, in a nutshell, comprise either an encoder or a decoder, or both, each stacked with multiple identical blocks. In general, the first block takes the tokenized sequence encoded by a word embedding layer, followed by a multi-head scaled-dot
self-attention (MHA) layer with an attention mask corresponding to specific language modeling objectives and a feed-forward network (FFN) layer. 
The attention mechanism~\cite{bahdanau2014neural} underpinning the Transformer architecture is implemented in the MHA layer, which computes a weighted representation of each token in the input sequence based on its relevance to others.
Slightly more technically, the word-embedded token sequence 
which normally concatenates long contexts and user prompts 
gives rise to three embedding matrices, i.e., 
the query $Q$, 
the key $K$ 
and the value $V$, 

on which the attention (kernel) operations are performed. 
\begin{gather}    
	P \defeq Q\times K^{\mathrm{T}}, \label{equa:qk}\\
	A \defeq \mathrm{softmax}[\cfrac{P}{\sqrt{d_k}}\odot M], \label{matrixA}\\
	O \defeq (A\times V) \times W_O, \label{equa:ov}
\end{gather}
Namely, assuming the token sequence length $L$,   
each entry of the (unnormalized) relevance matrix $P \in \domain{L \times L}$ measures the relevance of the corresponding pair of tokens. 
The normalized \emph{attention weight} matrix $A \in \domain{L \times L}$ is computed as a scaling operation 
and an element-wise mask operation with $M\in \domain{L\times L}$, together with a row-wise softmax. Finally, the output hidden state matrix $O$ 
is generated by a weighted sum of $V$ with attention weights in each row of $A$, usually with an extra linear transformation $W_O$. 

The attention mechanism is very costly, albeit effective. 
To reduce its computational demands, a common strategy is to use the Key-Value (KV) cache. In a nutshell, it is a list of tensors that stores the $K, V$ embeddings 
for all previous tokens in the attention layer for each block (prefilling), utilized and 
updated during the autoregressive generation process of 
LLMs (decoding). 
A deficiency of KV caching is that LLMs (with billions of parameters) may consume substantial additional memory during the decoding stage, as they need to cache extensive KV states~\cite{hong2023flashdecoding++,yue2024wkvquant,wang2024loma}. For instance, CodeLlama-7B~\cite{roziere2023code} (which requires 14 GB to store model parameters) needs an additional 16 GB for the KV cache, under a batch size of 32 and a sequence length of 1,024.\footnote{We assess the storage overhead using fp16 precision. In the case of fp32, the KV cache demands 32 GB.} 
The memory demand poses challenges for deploying these models, especially in low resource environments.

Various methods have been proposed to reduce the size of KV cache. For instance, window attention~\cite{chen2023extending, beltagy2020longformer} and StreamingLLM~\cite{xiao2023efficient} predict subsequent words by only caching the most recent KV states. H\textsubscript{2}O~\cite{zhang2024h2o} and FastGen~\cite{ge2023model} have explored to preserve subsets of the states crucial for prediction by employing specific patterns. 
In this paper, these methods are collectively referred to as \emph{KV compression methods}. To a large extent, they leverage sparse, low-rank attention approximation~\cite{chen2021scatterbrain}, based on the belief that a subset of tokens contributes the most values when performing attention operations.

Although current KV compression methods turn out to be fairly effective in NLP tasks (related to dialogue and text completion), it is risky to apply them in code generation. Fundamentally, these methods typically encourage models to focus on local information. 
In code generation tasks, however, excessive reliance on local information may result in discrepancies between the generated code snippet and either user's intention (e.g., in the prompt) or the ongoing decoding process (e.g., from the context). This primarily stems from the inherent complexity of code, which naturally exhibits long-range dependencies. For instance, in repository-level code generation~\cite{wang2024teaching,shrivastava2023repository,zhang2023repocoder}, the relevant context that needs to be considered during generation comes from not only the current, but also externals, files, e.g., imported packages, source code files in the same directory, configuration files and even API documentation. In many cases, these artifacts have their own dependencies. Capturing these long-range dependencies demands more than mere understanding of the local context. 

\begin{figure}[t]
	\centering
	\includegraphics[width=0.95\linewidth]{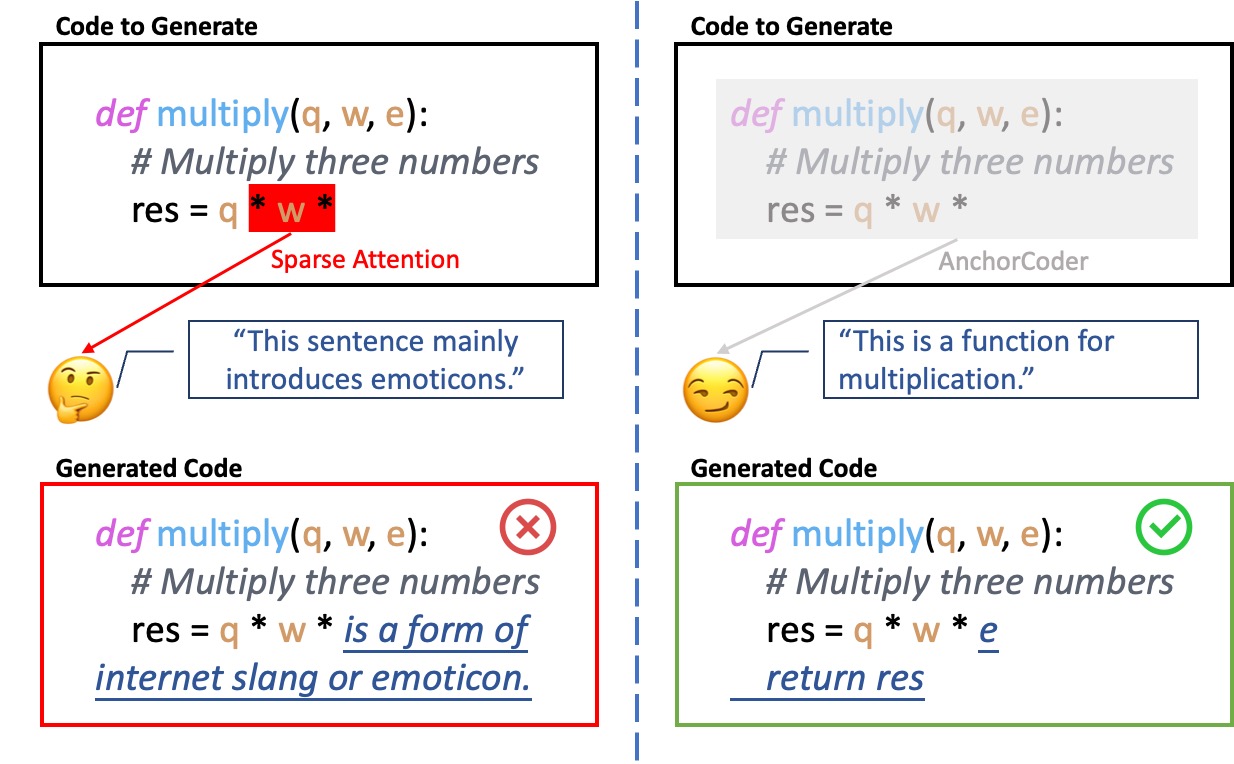}
	\caption{AnchorCoder's performance on context compression.
	}
	\label{fig:example}
\end{figure}

We present an example to illustrate the limitations in current KV compression methods 
in the left part of Fig.~\ref{fig:example}, where the LLM is supposed to generate code for multiplication of three numbers. 
The content within the sliding window (which captures the local information) only includes `* w *'. As a result, the model erroneously interprets this as 
symbolic emoticons, which wholly deviates from user's 
request (of a mathematical function).

In this paper, we aim to explore KV compression techniques devised for code LLMs without the over-reliance on local information. To this end, we first carry out an empirical study on code LLMs, based on which we introduce {\tool}, a novel approach that leverages ``anchors" to aggregate sufficient contextual information.   

\smallskip
\noindent\textbf{Empirical study.} To verify code LLMs' potential for KV compression, we first identify the sparsity pattern of the attention weights matrix $A$ in Eq.~(\ref{matrixA}) 
within code LLMs. 
We 
use the Gini coefficient~\cite{hurley2009comparing} and the sum of top-2 attention weights to measure the sparsity degree of the attention weights.

Our empirical study (cf.\ Section~\ref{sect:emp1}) reveals that code LLMs exhibit high sparsity on attention weights.
In the majority of layers, the Gini coefficient for attention weights exceeds 0.9. Furthermore, the sum of top-2 attention weights typically  accounts for 80\% of the total weights. This implies that, in these layers, the model concentrates on a subset of KV states to complete generation, while the vast majority of the KV cache is largely redundant.

Importantly, we discover that code LLMs exhibit a phenomenon of information aggregation
around specific tokens. 
These tokens, referred to as \emph{anchor points}~\cite{wang2023label,huang2024opera,zhang2024anchor,pang2024anchor}, are identified 
in the first layer which aggregate essential contextual information, particularly the semantics of each line of code. They 
enable the model to effectively summarize and distill the essential information for subsequent computation. As the computation progresses through later layers, a prominent concentration of attention weights on these anchor points is observed.

Furthermore, we evaluate the state-of-the-art KV compression methods to see whether they can capture the sufficient contextual information 
with different context lengths. To this end, we design a ``needle in a haystack'' experiment~\cite{kuratov2024babilong,chaudhury2024needle} tailored for code generation. 
The experiments reveal that, while the existing methods can achieve a high accuracy for shorter code snippets, their performance diminishes significantly for longer ones, 
where the ``needle'' is deeply embedded. 
In such cases, the model frequently fails to detect the ``needle'', and consequently, may not adhere to the given instructions. 
As illustrated in Fig.~\ref{fig:example}, the word `Multiply' is buried too deeply, leading the model to misinterpret the context due to the sparse attention mechanism. The details are given in Section~\ref{sec:needle}. 

\smallskip
\noindent\textbf{New approach.}
We present {\tool}, a novel approach 
designed to reduce storage demands of KV caches in code generation models while preserving essential contextual information. 
In a nutshell, {\tool} utilizes a mechanism that ``communicates in superposition"\cite{elhage2021mathematical}, aggregating the context to a few planted anchors. 
Let us revisit the example in Fig.~\ref{fig:example}.
Typical sparse attention mechanism 
tends to ignore context outside the sliding window (i.e., `* w *'). In contrast, {\tool}, as shown in the right part of Fig.~\ref{fig:example}, ensures effective code generation by compressing a sufficient context. 
The rationale lies in 
the compression phenomenon revealed in the empirical study, which
can reduce the size of context inherently, but does not substantially degrade model's performance. 

More technically, {\tool} features multi-head positional encoding (cf.\ Section~\ref{sec:token-wise}) and layer-wise anchor attention (cf.\ Section~\ref{sec:layer-wise}), which respectively address the loss of positional information due to compression and the degradation of information during transmission between layers. 

To evaluate the performance of {\tool}, we conduct experiments on three benchmark datasets, i.e., HumanEval~\cite{chen2021evaluating}, HumanEvalPlus~\cite{liu2024your} and MBPP~\cite{austin2021program}. The experimental results demonstrate that {\tool} maintains model performance at 102\%, 110\%, and 97\% on these three datasets, while achieving KV cache budget of 30\%, 30\% and 28\%, through efficient tuning. 
Furthermore, we design an experiment that trains {\tool} from scratch where the results show that with a KV cache budget of 30\%, {\tool} can still achieve performance comparable to that of dense attention, thereby validating the effectiveness and generalizability of {\tool}.

Our contributions can be summarized as follows.

\begin{itemize}
    \item We identify patterns of sparsity in the attention mechanisms of code LLMs and uncover the phenomenon of information aggregation on anchor points within them.  Additionally, we reveal the limitations of current KV compression methods on code LLMs.
    \item We propose {\tool}, a novel sparse attention based approach, which compresses context through token-wise anchor attention and mitigates information degradation through layer-wise anchor attention. This approach can reduce the KV cache overhead while preserving sufficient contextual information.
 
\end{itemize}

To the best of our knowledge, this is the first systematic research on effective KV compression methods in  LLMs for code generation, and software engineering in general. 

\smallskip\noindent\textbf{Organization.} The remainder of this paper is organized as follows. Section~\ref{sec:emp} presents an empirical study on code LLMs.
Section~\ref{sec:method} presents the proposed approach. 
Section~\ref{sec:setup} gives the experimental design and Section~\ref{sec:result} reports the results. Section~\ref{sec:threat} discusses potential threats to validity. Section~\ref{sec:related} reviews the related work. Section~\ref{sec:conc} concludes the paper.

The source code of {\tool} is available at \url{https://github.com/NUAAZXY/Anchor_Coder} and the models are available at \url{https://huggingface.co/AnchorCoder}.

%% file: sections/3.empirical.tex
\section{Empirical Study} \label{sec:emp}
In general, models with sparse attention weights are 
relatively easier to be compressed, as only a limited number of KV states are needed. It is crucial to study the sparsity pattern of attention weights in code LLMs which is largely uncharted.

\subsection{Sparsity Pattern of Code Generation Models} \label{sect:emp1}
We carry out an empirical study with four code LLMs of varying scales, i.e., CodeGPT-0.1B~\cite{lu2021codexglue}, PolyCoder-0.4B~\cite{xu2022systematic}, CodeGen-2B~\cite{nijkamp2022codegen}, CodeLlama-7B~\cite{roziere2023code}
on three typical benchmark datasets, i.e., HumanEval~\cite{chen2021evaluating}, HumanEvalPlus~\cite{liu2024your} and MBPP~\cite{austin2021program}. 
These datasets comprise programming challenges designed to assess functional correctness and  user prompts alongside corresponding code.

Given a vector $\vec{w} = (w_1, \cdots, w_n)$ of extracted attention weights, where $w_i$ denotes the attention weight on a specific token,
we consider two metrics, i.e., Gini coefficient and the sum of top-2 weights, to measure the sparsity of attention weights distribution. 
The Gini coefficient is a standard metric for assessing sparsity~\cite{hurley2009comparing}, which is given by $G = \frac{\sum_{i=1}^n \sum_{j=1}^n |w_i - w_j|}{2n^2\bar{w}}$, where $\bar{w}$ is the mean of all attention weights. 
The sum of top-2 attention weights, is defined as the sum of the two highest attention scores within $S$. Formally, it is given by $T = w_{(1)} + w_{(2)}$ where $w_{(1)}$ and $w_{(2)}$ denote the highest and second highest weight in $\vec{w}$. The sum of top-2 attention weights provides an intuitive measure of concentration within attention, indicating the proportion of attention weights attributed to the key positions that the model focuses on the most.

\begin{figure}[t]
	\centering
	\begin{minipage}[c]{0.95\linewidth}
		\includegraphics[width=\linewidth]{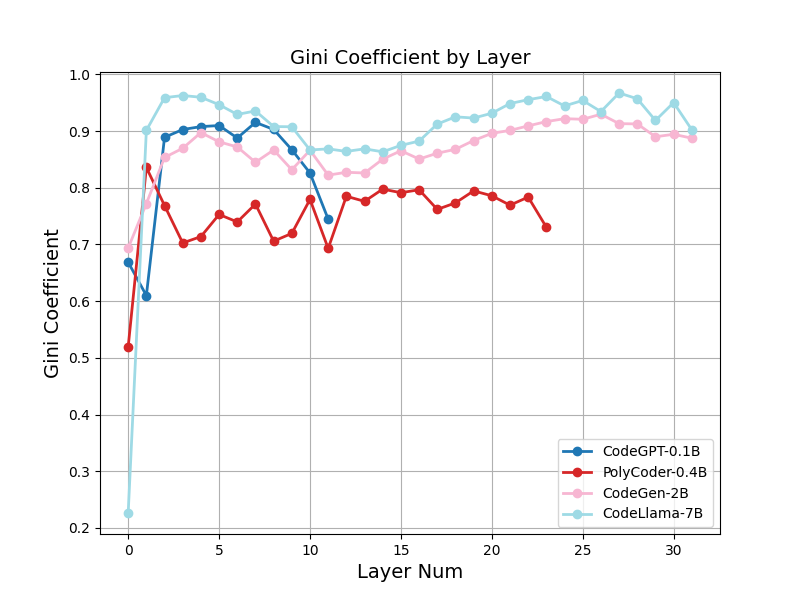}
	\end{minipage}
	\begin{minipage}[c]{0.95\linewidth}
		\includegraphics[width=\linewidth]{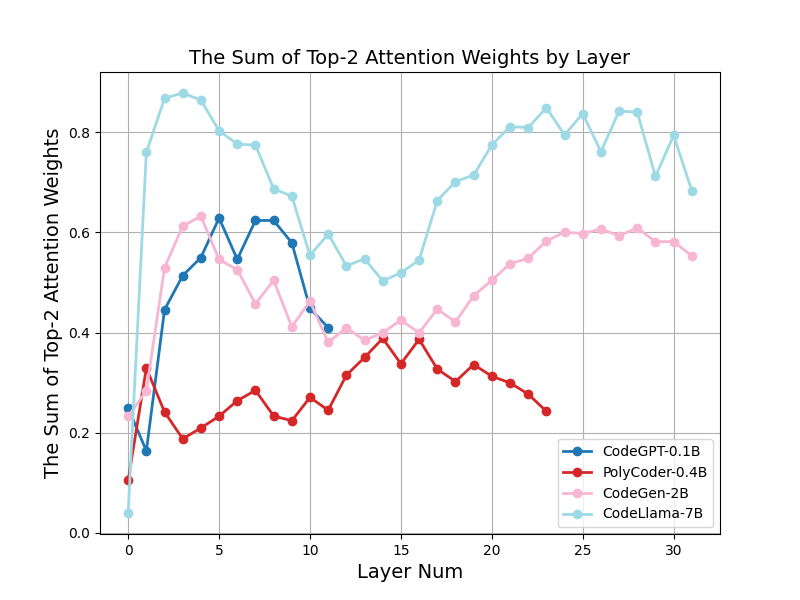}
	\end{minipage}
	\caption{Sparsity of attention weights in code LLMs.}
	\label{fig:sparse_ana}
\end{figure}

\begin{figure}[t]
	\centering
	\includegraphics[scale=0.375]{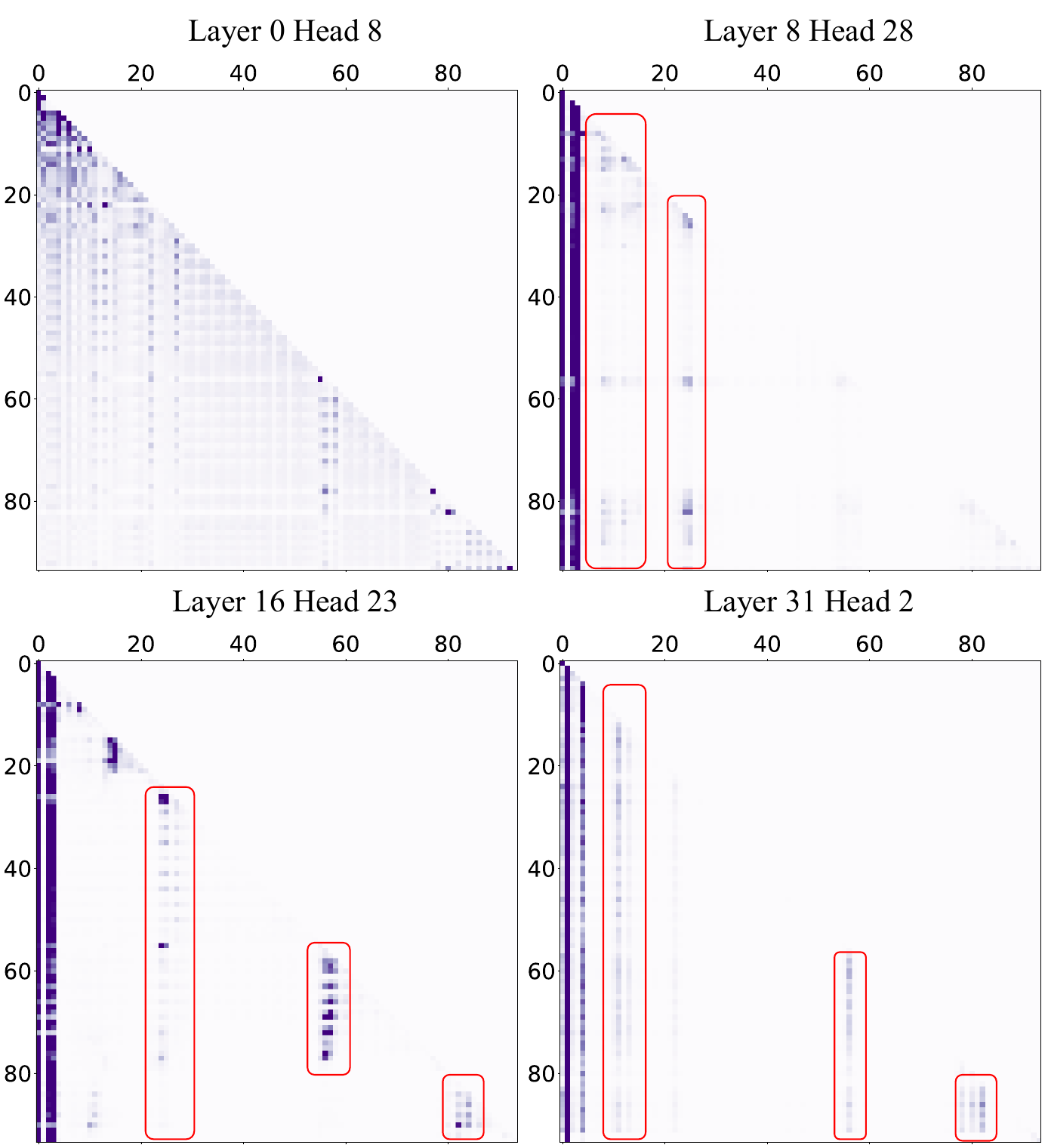}
	\caption{The attention heatmap of CodeLlama-7B.}
	\label{fig:attention}
\end{figure}


Fig.~\ref{fig:sparse_ana} presents the Gini coefficient and the sum of top-2 attention weights for each layer of the four models, where higher values (approaching 1) indicate higher sparsity.
Clearly, except for the first layer, the distributions of attention weights tend to be sparse. This indicates that in the majority of layers within the model, attention is predominantly concentrated on a small subset of KV states, while the attention weights on other positions are close to zero\footnote{In this paper, `KV state' and `position' refer to location in a particular layer, while `token' refers to location across all layers.}.
Notably, larger models, especially those with increased dimensions in their hidden states, tend to achieve a higher degree of sparsity. 
For instance, CodeLlama-7B shows the highest 
Gini coefficient and sum of top-2 attention weights.

Importantly, the higher sparsity exhibited by the attention weights does \emph{not} imply that the model 
can work based on only few tokens. Rather, it suggests that the positions with high attention weights aggregate the contextual information. 
The attention mechanism operates by computing attention weights through the dot product of previous query states and key states (Eq.~\ref{equa:qk}), which is further used to generate the hidden state via a weighted sum of value states (Eq.~\ref{equa:ov}). 
This indicates that 
these positions 
encode content from previous contexts rather than just 
a particular token.

We also present an attention heatmap for CodeLlama-7B, depicted in Fig.~\ref{fig:attention} which visualizes the attention weights between tokens. (The darker colors represent higher attention weights.)
In Layer 0, we notice that the attention weights are densely distributed, with the model allocating relatively even attention to each token.
In contrast, at deeper layers (e.g., the 8th, 16th and 31st layer), the model tends to focus on fewer pieces of aggregated information, 
as highlighted by the red boxes.
As mentioned in Section~\ref{sect:intro}, the specific tokens that the model concentrates on are referred to as anchor points, 
where the model aggregates and summarizes previous  information in the initial layer via dense attention. As the processing proceeds to deeper layers, the model then uses these anchor points to predict the next token.  
In addition, the first few tokens in the code receive very high attention weights and often represent absolute positions, known as `sink tokens', as introduced in~\cite{xiao2023efficient}

To delve deeper into the patterns of attention weights in code LLMs, we examine the tokens (excluding `sink tokens') that garnered the most attention across the three datasets (i.e., HumanEval, HumanEvalPlus and MBPP), as shown in Table~\ref{tab:attention}. (In this table, \emph{num} represents the frequency of these tokens receiving the highest attention weights, while \emph{ratio} indicates their proportion relative to the total number of tokens analyzed.) 
Surprisingly, the model does not predominantly focus on tokens that carry critical semantic content, such as Python keywords. Instead, it predominantly focuses on relatively semantic-free tokens such as  linebreak tokens (`\textbackslash n'). 
For instance, in CodeLlama, 78.2\% of the attention is concentrated on `\textbackslash n', while only 21.8\% is distributed among other tokens.
This further illustrates the phenomenon of information aggregation within the model, as 
these tokens would be inadequate for prediction. The possible explanation would be that the model compresses contextual information into these `\textbackslash n', highlighting a unique aspect of compression mechanism in code LLMs.

In summary, by a thorough analysis of the attention mechanism for representative LLMs for code generation, we confirm the sparsity pattern of the model's attention and find that the linebreak token acts as an anchor point in the model, enabling the compression of each line of code. This discovery highlights the tremendous potential of using sparse attention for code generation by leveraging the compressed information.

\begin{table}[htbp]
	\centering
	\caption{Distribution of Attention Weights on Tokens}
	\begin{tabular}{lcccc}
		\hline
		\multirow{2}{*}{Model} & \multicolumn{2}{c}{`\textbackslash n'} & \multicolumn{2}{c}{others} \\ 
		& num & ratio & num & ratio  \\ \hline
		CodeGPT-0.1B & $1.5 \times 10^5$ & $24.8\%$ & $4.6 \times 10^5$ & $75.2\%$ \\ 
		PolyCoder-0.4B & $3.5 \times 10^5$ & $32.4\%$ & $7.2 \times 10^5$ & $67.6\%$ \\ 
		CodeGen-2B & $1.4 \times 10^6$ & $90.8\%$ & $1.4 \times 10^5$ & $9.2\%$ \\ 
		CodeLlama-7B & $1.3 \times 10^6$ & $78.2\%$ & $3.5 \times 10^5$ & $21.8\%$  \\ \hline
	\end{tabular} 
	\label{tab:attention}
\end{table}


\subsection{Evaluation of Existing KV Compression Methods} \label{sec:needle}
As the sparsity of attention weights within code LLMs is confirmed, 
a natural question is whether the state-of-the-art methods, (such as StreamingLLM~\cite{xiao2023efficient} and H\textsubscript{2}O~\cite{zhang2024h2o}) in NLP can be applied to code LLMs directly. 
To answer this question, 
we design a ``needle in a haystack'' experiment~\cite{kuratov2024babilong,chaudhury2024needle} tailored for code generation. 

We construct a list  \code{l=[$x_{1}, x_{2}, \cdots , \text{'needle'},\cdots, x_{n}$]} of length $n$, where $x_i$'s are randomly generated numbers, 
and instruct LLMs to complete the code \code{assert 'needle' in l ==} and \code{assert 'needle' not in l ==} with the prompt ``\# Determine if there is `needle' in this list, and complete the code by filling in True or False.'' We vary $n$ from 64 to 512, with the `needle' uniformly distributed across positions within the list.

The purpose of this experiment is to observe whether a model using the sparse attention mechanism can locate the ``needle'', reflecting their capability in retrieving contextual information. 
As it is a binary classification task, the prediction accuracy should ideally be significantly greater than 0.5; otherwise 
it suggests that the model fails to follow instructions 
to complete the code. 
Our designed experiment provides a direct assessment of the model's information context retrieval capability and its adherence to prompts~\cite{liu2024exploring}. Compressing the KV cache could potentially weaken these abilities, which are crucial for generating correct code.

\begin{figure}[t]
    \centering
    \begin{minipage}[c]{0.9\linewidth}
        \includegraphics[width=\linewidth]{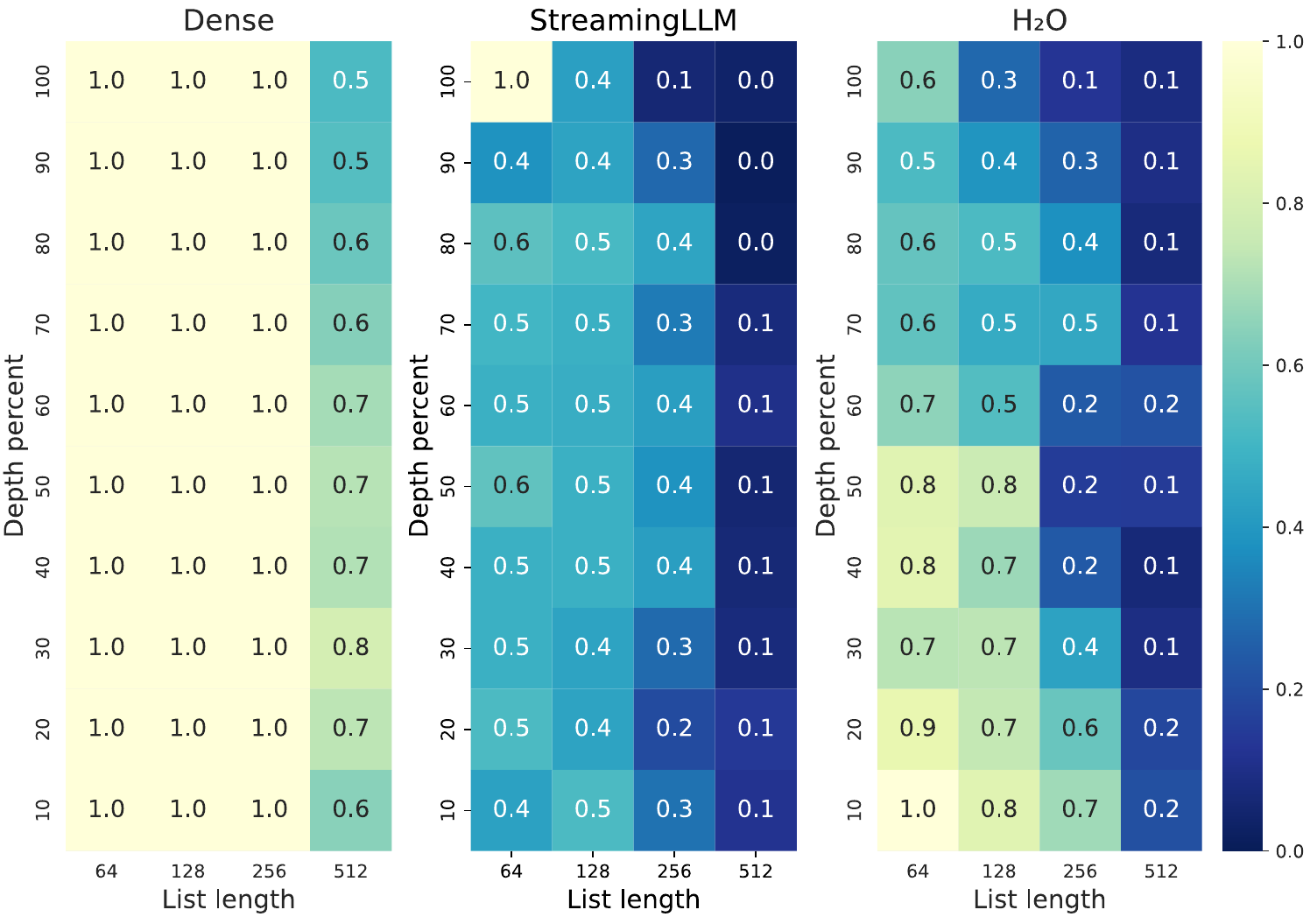}
    \end{minipage}
    \caption{The results of the ``needle-in-a-haystack" experiment.}
    \label{fig:needle}
\end{figure}

The results, illustrated in the Fig.~\ref{fig:needle}, show a comparison between sparse and dense attention methods across varying context lengths and depths of ``needle'' placement. StreamingLLM and H\textsubscript{2}O display  high accuracies with shorter list lengths, indicating that limited attention extraction can handle context information and follow prompts in short texts. However, as the list length increases (with deeper ``needle'' positions), 
the performance of these models experiences a significant degradation 
with accuracy approaching zero, primarily due to the constraints imposed by the window size which prevents the models from querying the input prompt effectively.
We hence can conclude that the method of extracting parts of the context using sparse attention 
is insufficient for code generation models.

%% file: sections/4.method.tex
\section{The {\tool} Method}\label{sec:method}
 
In this section, we propose {\tool}, a novel KV compression method. We first introduce the token-wise anchor attention mechanism to reduce the required size of the KV cache (Section~\ref{sec:token-wise}), followed by the layer-wise anchor attention to mitigate the superposition problem  (Section~\ref{sec:layer-wise}).

\subsection{Token-wise Anchor Attention}\label{sec:token-wise}
\begin{figure*}[t]
	\centering
	\includegraphics[width=0.95\linewidth]{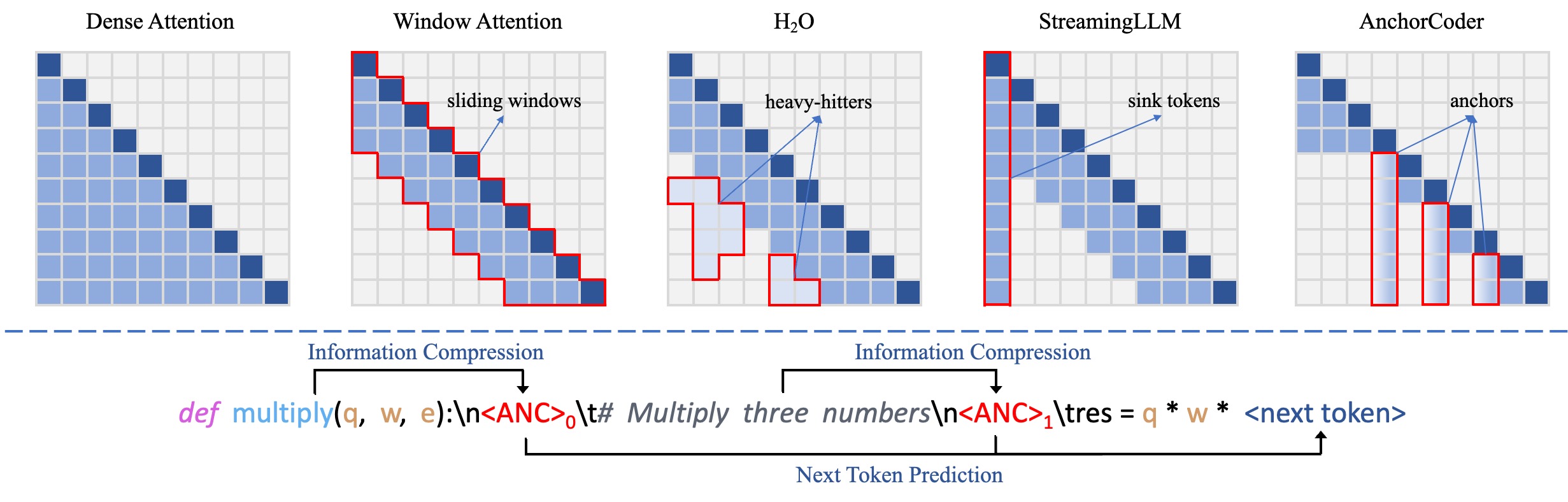}
	\vspace{-0.2cm}
	\caption{The upper part provides the attention heatmap deploying different KV cache compression methods. The lower part provides the illustration of {\tool}.
	}
	\label{fig:workflow}
\end{figure*}

The empirical study in Section~\ref{sec:emp} reveals that the distribution of attention weights in code generation models is highly sparse and tends to aggregate information on certain anchor points. 
Although most of these anchor points are linebreak tokens (e.g., making up 78.2\% in CodeLlama-7B), relying solely on them may lead to contextual information loss (21.8\% is discarded). To address this issues, 
we introduce \underline{T}oken-wise \underline{A}nchor \underline{A}ttention (TAA), 
a method that plants artificial anchors for each line of code and trains them as aggregator  
of contextual information. 
An illustration of {\tool} is given in Fig.~\ref{fig:workflow} (lower part). 
Compared to Window Attention and its variants, e.g., H\textsubscript{2}O (which includes `heavy-hitters') and StreamingLLM (which adds `sink tokens'), 
{\tool} 
plants the anchor $\textless \text{ANC}\textgreater_0$ to compress the method name and parameters, and $\textless \text{ANC}\textgreater_1$ 
to compress the comments on function's purpose as shown in Fig.~\ref{fig:workflow}.
By these predefined anchors, the model can perform attention operations merely on these positions, 
reducing the required number of KV states. 

Importantly, {\tool} preserves the original attention sparsity patterns of code LLMs, ensuring that their core functionalities remain unaffected. A comparison of the heatmaps from {\tool} in Fig.~\ref{fig:workflow} (upper part) with those from dense attention mechanisms in Fig.~\ref{fig:attention} reveals a strong similarity which highlights {\tool}'s capability to effectively leverage existing attention behaviors. 
Noteworthy, the similarity in attention patterns enables the model to be tuned efficiently without requiring extensive computational resources.
Indeed, we employ low-rank adaptation (LoRA)~\cite{hu2021lora}, which achieves training efficacy by tuning a small set of parameters.
	
Formally, assume a code snippet (from a training dataset) takes the following form, 
$$D = x_1, \ldots, x_p, \backslash n, x_{p+1}, \ldots, x_q, \backslash n, \ldots, x_n$$ which comprises multiple lines of code 
separated by linebreak tokens `\textbackslash n'. (Here, the total length of the code snippet is $n$ and each $x_i$ is a token.) 

We append the special token {\textless ANC\textgreater}\footnote{Note that {\textless ANC\textgreater}  is only used in the computational and will not appear in the generated code.} after each \textbackslash n,
obtaining $D_{\text{anchored}} =$
$$x_1, \ldots, x_p, \backslash n, \textless \text{ANC}\textgreater_0, x_{p+1}, \ldots, x_q, \backslash n, \textless \text{ANC}\textgreater_1,  \ldots, x_n$$

To restrict model's concentration on anchors, we compute an attention mask (as per $M$ in Eq.~(\ref{matrixA})) in Algorithm~\ref{algo:att}. This attention mask $M$ comprises two elements, i.e., autoregressive mask (Line 3-4) and anchor mask (Line 5-9). The autoregressive mask is designed to prevent the model from attending to future tokens, which is achieved by setting the weights of these positions to a very small value, i.e., $neg\_inf$, making their attention weights zero after applying softmax.
The anchor mask is designed to mask tokens between anchors. Specifically, given a set $\mathcal{I}$ of anchor indices (e.g., in $D_{anchored}$ the anchor index for $\textless \text{ANC}\textgreater_0$ is $p+2$), we mask all tokens located between two consecutive anchor indices to ensure that attention is concentrated solely on these anchors.

\begin{algorithm}[ht]
	\footnotesize
	\caption{Attention Mask Algorithm} 
	\label{algo:att}
	\KwIn{
		\\
		Input Length $n$ and Anchor Indices $\mathcal{I}$;\\
	}
	\KwOut{
		\\
		Attention Mask $M$;\\
	}
	$M \gets 0_{n \times n}$ \tcp*{Initialize $M$ as an $n \times n$ zero matrix}
	$neg\_inf = -1e9$ \tcp*{A Small Number Ensuring the Result After Applying Softmax is Close to Zero.}
	\For{$i = 0$ \KwTo $n-1$}{
		$M[i, i+1:n] = neg\_inf$ \tcp*{Add Autoregressive Mask by Set Future Tokens Inviable}
		\For{$j = 0$ \KwTo $|\mathcal{I}|-2$}{
			\If{$\mathcal{I}[j+1] < i$}{
				$Start \gets \mathcal{I}[j]$; \\
				$End \gets \mathcal{I}[j+1]$; \\
				$M[i, Start:End] = neg\_inf$\tcp*{Add Anchor Mask by Set Tokens between Anchors Inviable}
			}
			\Else{
				\Break
			}
		}
	}
	\Return $M$
\end{algorithm}

Our context compression approach minimizes information loss through ``communication in superposition", effectively preserving the semantics of the context.
However, 
in its plain form, it 
primarily considers the positional information of anchors when calculating attention weights, 
disregarding the relative positions of overlapping contexts within them.
In code LLMs, positional information is crucial~\cite{peng2022rethinking}, as even slight deviations can significantly alter the semantics.

To address this issue, inspired by the concept of Multi-Head Attention (MHA), we propose \underline{M}ulti-\underline{H}ead \underline{P}ositional \underline{E}ncoding (MHPE), which incorporates independent positional encodings into the various attention heads of anchors, thereby mitigating the loss of positional information.

In MHA, model's query, key, and value vectors are divided into multiple subspaces. Each attention head processes information from a distinct subspace, focusing on a specific subset of tokens and capturing their unique features and relationships within that subspace. This division allows the model to attend to different tokens across various attention heads, thereby enhancing its ability to process complex contextual information.

When handling tokens compressed within anchors, an optimal strategy is to assign individual positional information to each token. To achieve this, we leverage the characteristic of MHA for its ability to extract features from different subspaces. We treat these compressed tokens as distinct features stored in the anchors, distributed across their respective subspaces. By computing feature correlations within these subspaces, we can effectively perform attention calculations on these compressed tokens.

To implement MHPE, we utilize \underline{Ro}tary \underline{P}osition \underline{E}mbedding (RoPE) as the positional embedding method. RoPE 
rotates the input vectors to 
supervise dependencies among tokens at varying positions within the sequence.  For anchors in $D_{\text{anchored}}$, we use specialized attention heads to process their subspaces independently, and incorporate positional information of the tokens compressed within these anchors. 
In detail, MHPE can be defined as 
\[ 
	\tilde{k} = \text{concat}(R_{s_0}k_0, \cdots, R_{s_{head}}k_{head}),  
\]
where for each $s_i\in S$,  $R_{\Theta,s_{i}}$ is a block diagonal matrix with blocks of the form 
\[
	(R_{s_i})_t = 
\begin{pmatrix}
	\cos s_i\theta_t & -\sin s_i\theta_t  \\
	\sin s_i\theta_t & \cos s_i\theta_t  \\
\end{pmatrix}
,
\quad
\theta_t=\theta^{-2t/d}
\]
for $t=1, \cdots, \frac{d}{2}$. 
Note that $\{k_0,\cdots, k_{head}\}$ represents a set of key states for attention heads with $d$ dimensions, and $S$ is a set of position indices between anchors.


\subsection{Layer-wise Anchor Attention}\label{sec:layer-wise}

In LLMs, residual stream plays a crucial role in passing information across layers. However, in {\tool}, the residual stream may be affected by its bottleneck, leading to information degradation.  The residual stream adds each layer's output back to its input before passing it to the next layer, ensuring a consistent flow of information. 
Intuitively, 
this reinforces the information from the previous layers, effectively allowing the model to track changes to the input as it propagates through the layers~\cite{elhage2021mathematical,Kawasaki2024DefendingLL}. Formally
\begin{align*}
	\hat{r}_l & =  r_{l-1} + \text{attn}(\text{LN}(r_{l-1})) \\
	     r_l  & = \hat{r}_l + \text{mlp}(\text{LN}(\hat{r}_l)), 
\end{align*}
where
\begin{equation}\label{eq:residual} 
\text{attn}(x) = W_O \times V \times \text{softmax}(\frac{Q \times K^T}{\sqrt{d_k}})
\end{equation}
Here, $r_l$ represents the state of the residual stream after writing information at the $l$-th layer, LN$(\cdot)$ denotes layer normalization,
mlp$(\cdot)$ denotes the multi-layer perceptron in transformer, $\hat{r}_l$ represents the intermediate representation in residual computation, and $W_Q, W_K, W_V \in \mathbb{R}^{d \times d}$ are matrices that transform the inputs to query, key and value vectors, respectively, with $d$ being the dimension of the model. 
Since each transformer layer writes information from the current layer to the residual stream, it often faces so called activation bottlenecks~\cite{elhage2021mathematical}, where the dimension required to write complete information from each layer is much higher than that of the residual stream itself.
As a result, it is virtually impossible to retain the complete information written by each layer, and the model must find ways to manage the limited capacity of the residual stream. Previous research has shown that attention heads 
writing in the opposite direction 
to the residual stream may effectively eliminate certain features and alleviate the problem of activation bottlenecks~\cite{dao2023adversarial}.

To determine whether a similar phenomenon happens in code LLMs, we analyze the eigenvalues of the matrix $W_{OV}=W_O \times W_V$. As per Eq.~\ref{eq:residual}, $W_{OV}$ writes ``linearly" 
to the residual stream and does not mix information between tokens, thereby directly influences the state updates within this stream. In contrast, $W_{QK} = W_Q \times W_K$ mixes information between tokens and is gated by the (nonlinear) softmax
~\cite{millidge2022singular}.
The presence of negative eigenvalues in $W_{OV}$ of attention heads would indicate that the model is removing information from the residual stream, revise{which may 
degrade the performance of the neural network. This is because negative eigenvalues suggest that certain dimensions in the feature space are being shrunk or collapsed, rather than being preserved or enhanced as intended. In the context of attention mechanisms, which are designed to 
focus on the most relevant parts of the input data, this could mean that the model is inadvertently ignoring or downplaying important features.}

\begin{figure}[t]
	\centering
	\includegraphics[width=0.9\linewidth]{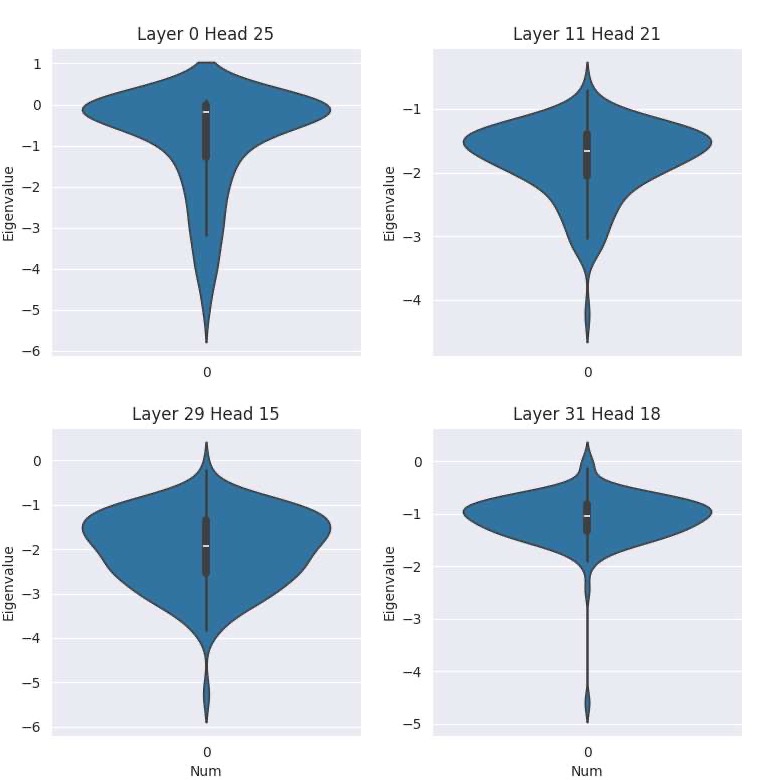}
	\vspace{-0.2cm}
	\caption{Eigenvalue distribution of $W_{OV}$.
	}
	\label{fig:WOV}
\end{figure}

Our analysis (Fig.~\ref{fig:WOV}) reveals that, in some attention heads within the model, a majority of the eigenvalues of $W_{OV}$ are negative. This suggests that these attention heads are writing in the opposite direction into the residual stream, deleting information and alleviating the activation bottleneck issue~\cite{elhage2021mathematical}.

Previous studies observed that models can make accurate predictions at shallow layers, which may become incorrect in deeper layers~\cite{chuang2023dola,sun2024neural}. This suggests that crucial evidence present in the shallow layers is being removed or discarded as information flows through the model. Logically, certain attention heads are responsible for deleting the evidence via the residual stream, leading to information degradation.
In {\tool}, 
as the anchors 
aggregate contextual information resulting in a superposition, 
the deletion hehavior of the attention heads may be amplified, thereby increasing the risk of 
losing crucial information in generation.

\begin{figure}[htbp]
	\centering
	\includegraphics[width=0.8\linewidth]{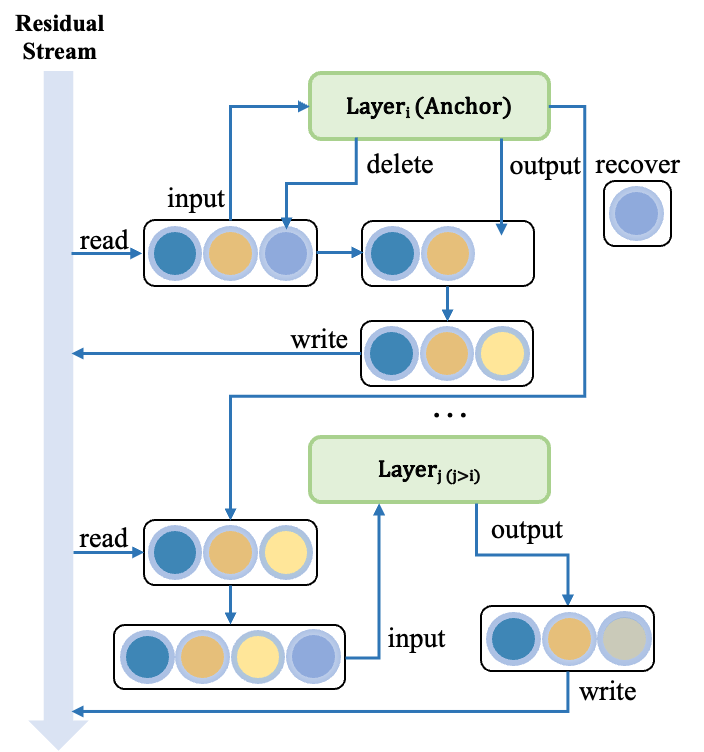}
	\caption{Workflow of layer-wise anchor attention.
	}
	\label{fig:LAA}
\end{figure}
To mitigate this issue, we introduce a simple yet efficient method, namely, \underline{L}ayer-wise \underline{A}nchor \underline{A}ttention (LAA), illustrated in Fig.~\ref{fig:LAA}. LAA alleviates the impact of information loss by setting a bypass for the residual stream. Specifically,
we set up an specific anchor layer, 
within the shallow layers of the model. 
The KV states of this anchor layer are then utilized as additional targets 
for attention calculations in deeper layers, which can be formalized as 
\begin{gather*}
	\text{LAA}(Q, K, V) = \text{softmax}\left(\frac{Q \times \tilde{K^T}}{\sqrt{d_k}}\right)\tilde{V}, \\
	\tilde{K} = \text{concat}(K, K'), 
	\tilde{V} = \text{concat}(V, V'),
\end{gather*}
where $Q, K, V$ are query, key and value states of the current layer, $K', V'$ denote key and value states of the anchor layer, $d_k$ represents the dimension of the model.

By computing attention scores that focus on these KV states, our approach effectively recovers crucial information that might otherwise be lost as data flows through the model layers. 
This design enhances the completeness of the residual stream, significantly boosting model's capacity to recognize and process shallow features. Notably, since LAA reuses the KV cache in anchor layer to recover information, it does not incur additional memory overhead. Moreover, although LAA requires extra computation, it does not significantly increase the inference latency, which will be discussed in Section~\ref{sec:ablation}.

%% file: sections/5.setup.tex
\section{Experiment Setup}\label{sec:setup}
We carry out extensive experiments for {\tool}  to address the following three research questions:

\begin{description}
\item[\textbf{RQ1:}] How accurate is the code generated by {\tool}?

\item[\textbf{RQ2:}] What are the KV cache requirements of {\tool}?

\item[\textbf{RQ3:}] How do MHPE and LAA contribute to model's performance?
\end{description}

\smallskip
\noindent\emph{Base Model.}
We use CodeLlama-7B~\cite{roziere2023code} (built upon LLaMA 2~\cite{touvron2023llama}) as the base model for {\tool}, which is referred to as ${\tool}_T$ henceforth. CodeLlama has been fine-tuned for ${\tool}_T$ using a modest dataset and low-rank approximation techniques to demonstrate the efficiency of fine-tuning ${\tool}$. CodeLlama-7B consists of 32 transformer blocks, each with 32 attention heads and a dimensionality of 4,096.
To mitigate the impact of utilizing additional data, we trained a Llama-like model from scratch for both the baseline and {\tool}. This model features 100 million parameters and a hidden size of 512, encompassing 16 transformer blocks and 8 attention heads. This variant of {\tool} is referred to as ${\tool}_P$. Since ${\tool}_P$ is initialized randomly, its attention remains dense, allowing us to assess the generalizability of {\tool} when applied to models with dense attention.

\smallskip
\noindent\emph{Datasets.}
We incrementally tune ${\tool}_T$ on the CodeSearchNet and CodeHarmony datasets, respectively, and evaluate the experimental results on the HumanEval~\cite{chen2021evaluating}, HumanEvalPlus~\cite{liu2024your} and MBPP~\cite{austin2021program} datasets. 
For ${\tool}_P$, we train from scratch using CodeSearchNet with a total of 500M tokens.

\begin{itemize}[leftmargin=*]
\item \textbf{CodeSearchNet:} The CodeSearchNet corpus~\cite{liu2023graphsearchnet} is a dataset comprising 2 million (comment, code) pairs from open-source libraries hosted on GitHub. It contains code and documentation for several programming languages. We utilize the Python code subset for incremental pre-training.

\item \textbf{CodeHarmony:} CodeHarmony\footnote{https://huggingface.co/datasets/Flab-Pruner/CodeHarmony} is 
curated from existing open-source datasets, and employs LLMs for automated test case generation. 

\item \textbf{HumanEval:} The HumanEval dataset~\cite{chen2021evaluating}, released by OpenAI, includes 164 programming problems with function signatures, docstrings, bodies, and several unit tests. 

\item \textbf{HumanEvalPlus:} The HumanEvalPlus dataset~\cite{liu2024your}, building upon the foundation established by HumanEval, has expanded its test cases by 80 times, thereby enhancing its capability to assess the correctness of code.

\item \textbf{MBPP:} The MBPP benchmark~\cite{austin2021program} consists of approximately 1,000 crowd-sourced Python programming problems, designed to be solvable by entry-level programmers. Each problem includes a task description, code solution and three automated test cases.
\end{itemize}

\smallskip
\noindent\emph{Baseline.}
%
To compare with our approach, we select \emph{Window Attention}~\cite{beltagy2020longformer}, \emph{StreamingLLM}~\cite{xiao2023efficient} and \emph{H\textsubscript{2}O}~\cite{zhang2024h2o} which are training-free methods based on sparse attention. In general, these methods  employ the concept of sliding window to reduce the number of tokens that the model attends to, and incorporate additional global information, such as attention sinks for StreamingLLM and heavy hitters for H\textsubscript{2}O.
In addition to these approaches, we also consider  training-based \emph{AutoCompressors}~\cite{chevalier2023adapting} as a baseline, which incorporates additional summary tokens for context compression and utilizes them as soft prompts for subsequent generation.

\smallskip
\noindent\emph{Metrics.}
To evaluate the performance of ${\tool}_P$, we employ pass@k and KV Cache Budget
as the metrics for evaluating model's performance. The pass@k metric reports the percentage of problems solved within the k generated codes. In this study, we utilize the pass@1 metric with greedy decoding for both {\tool} and baselines, which provides the most direct reflection of the model's generative capability and is not subject to the influence of randomness. 
Considering the different strategies employed by various methods, we are unable to set the same budget for each model. Therefore, we have set a similar KV cache budget for each model, allowing for a 1.5\% deviation.

Due to the limited computational resource, we only use 500M tokens to train ${\tool}_P$, which is challenging for generating correct code in HumanEval and MBPP dataset. Therefore, we use perplexity and accuracy metrics to evaluate the language modeling and next token prediction capabilities of different methods respectively.

\smallskip
\noindent\emph{Implementation Detail.}
For ${\tool}_T$ and its corresponding baseline, we tuned CodeLlama-7B on the CodeSearchNet and CodeHarmony dataset with 150M tokens, for which we use Low-Rank Adaptation (LoRA)~\cite{hu2021lora} to fine-tune the $W_Q$, $W_K$, $W_O$, and $W_V$ matrices of the models with rank 16 and learning rate 5e-5. The training of ${\tool}_T$ is  done on a single Nvidia RTX 4090. We control the sparsity of the attention weights by creating three variants, each with a different number of layers utilizing TAA. Specifically, we configure the models to use TAA in 24, 28, and 30 out of 32 layers, with the remaining layers adopting dense attention to aggregate context. The anchor layers for the three variants are respectively positioned at the 8th, the 8th layer, and the 16th layer.

For ${\tool}_P$ and its corresponding baseline, we employ  the 500M tokens from CodeSearchNet dataset for full-parameter training with a learning rate of 5e-4. This training was accomplished using 5 Nvidia RTX 4090, with 12 out of 16 layers utilizing TAA, while the remaining layers employed dense attention.

All experimental results are produced using greedy decoding to eliminate randomness.

%% file: sections/6.result.tex
\begin{table*}[htbp]
	\centering
	\caption{Performance Comparison of ${\tool}_T$ in RQ1}
	\begin{tabular}{lcccccc}
		\hline
		\multirow{2}{*}{Method} & \multicolumn{2}{c}{HumanEval} & \multicolumn{2}{c}{HumanEvalPlus}& \multicolumn{2}{c}{MBPP} \\
		\cline{2-7}
		& KV Budget \% ($\downarrow$)&Pass@1 \% ($\uparrow$) & KV Budget \% ($\downarrow$)& Pass@1 \%  ($\uparrow$)& KV Budget  \% ($\downarrow$)&Pass@1 \% ($\uparrow$) \\
		\hline
		\multicolumn{7}{c}{\textbf{Oracle}} \\
		\hline
		Dense&100&31.10&100&23.17&100&39.40\\
		\hline
		\multicolumn{7}{c}{\textbf{Training-free methods}} \\
		\hline
		\multirow{3}{*}{Window Attention} &30&0.00&30&0.00&28&0.00\\
		&\cellcolor{gray!30}20&\cellcolor{gray!30}0.00&\cellcolor{gray!30}20&\cellcolor{gray!30}0.00&\cellcolor{gray!30}18&\cellcolor{gray!30}0.00\\
		&\cellcolor{gray!50}15&\cellcolor{gray!50}0.00&\cellcolor{gray!50}15&\cellcolor{gray!50}0.00&\cellcolor{gray!50}12&\cellcolor{gray!50}0.00\\
		\multirow{3}{*}{StreamingLLM} &30&6.10&30&4.27&28&3.40\\
		&\cellcolor{gray!30}20&\cellcolor{gray!30}1.83&\cellcolor{gray!30}20&\cellcolor{gray!30}1.22&\cellcolor{gray!30}18&\cellcolor{gray!30}1.60\\
		&\cellcolor{gray!50}15&\cellcolor{gray!50}1.22&\cellcolor{gray!50}15&\cellcolor{gray!50}0.00&\cellcolor{gray!50}18&\cellcolor{gray!50}1.80\\
		\multirow{3}{*}{H\textsubscript{2}O} &30&17.68&30&14.02&28&17.60\\
		&\cellcolor{gray!30}20&\cellcolor{gray!30}14.63&\cellcolor{gray!30}20&\cellcolor{gray!30}10.98&\cellcolor{gray!30}18&\cellcolor{gray!30}15.40\\
		&\cellcolor{gray!50}15&\cellcolor{gray!50}14.02&\cellcolor{gray!50}15&\cellcolor{gray!50}10.36&\cellcolor{gray!50}12&\cellcolor{gray!50}14.00\\
		\hline
		\multicolumn{7}{c}{\textbf{Training-based Methods}} \\
		\hline
		\multirow{3}{*}{AutoCompressors} &30&24.39&30&21.34&28&23.00\\
		&\cellcolor{gray!30}20&\cellcolor{gray!30}23.17&\cellcolor{gray!30}20&\cellcolor{gray!30}20.12&\cellcolor{gray!30}18&\cellcolor{gray!30}20.20\\
		&\cellcolor{gray!50}15&\cellcolor{gray!50}17.07&\cellcolor{gray!50}15&\cellcolor{gray!50}13.41&\cellcolor{gray!50}12&\cellcolor{gray!50}17.00\\
		\multirow{3}{*}{${\tool}_T$} &30&\textbf{31.71}&30&\textbf{25.61}&28&\textbf{38.20}\\
		&\cellcolor{gray!30}20&\cellcolor{gray!30}\textbf{29.88}&\cellcolor{gray!30}20&\cellcolor{gray!30}\textbf{25.00}&\cellcolor{gray!30}18&\cellcolor{gray!30}\textbf{38.00}\\
		&\cellcolor{gray!50}15&\cellcolor{gray!50}\textbf{27.44}&\cellcolor{gray!50}15&\cellcolor{gray!50}\textbf{24.3}&\cellcolor{gray!50}12&\cellcolor{gray!50}\textbf{36.40}\\
		\hline
	\end{tabular}
	\label{tab:baseline}
\end{table*}

\section{Experiment Results}\label{sec:result}

\subsection{RQ1: Accuracy of the generated code}

To address \textbf{RQ1}, we evaluate {\tool} on the HumanEval, HumanEvalPlus and MBPP datasets, respectively. 
To ensure fairness, we set the sparsity in the baseline to the same level.
Notably, due to the varying compression strategies employed by each method, the length of the generated code differs. Consequently, it is challenging to set the sparsity level exactly the same. We control the sparsity difference between methods to be within 1.5\%.

The experimental results, shown in Table~\ref{tab:baseline}, indicate that {\tool} surpasses the baselines and achieves performance comparable to dense attention with KV cache budget at 30\%.

Local attention-based methods such as Window Attention and StreamingLLM tend to generate code with hallucinations and inconsistent with the prompt, due to their inherent limitation in concentrating only on local information and failing to capture the user's intent specified in the prompt. 
Additionally, Window Attention cannot correctly generate tokens beyond the window size because it overlooks the tokens at the beginning of the text, which are crucial for representing the absolute position in the text~\cite{xiao2023efficient}.

H\textsubscript{2}O achieves attention compression by discarding tokens with low attention weights in previous computations, assuming these tokens are insignificant for predicting subsequent tokens. However, in code generation tasks the spatial positions of related code fragments can be far apart, and H\textsubscript{2}O might discard tokens that are insignificant in the earlier context but critical in the subsequent context. Moreover, AutoCompressors uses multiple summary tokens as soft prompts to represent more contextual information with a few key-value caches. However, it 
assigns absolute positional information to these tokens, potentially affecting model's generalization ability. Furthermore, the summarization method based on soft prompts can disrupt the positional information of the context. In contrast, {\tool} compresses context into the anchor points, thereby maintaining the model's generation capability with less KV states. 

\begin{table}[htbp]
	\centering
	\caption{Performance Comparison of ${\tool}_P$ in RQ1}
	\begin{tabular}{lcc}
		\hline
		Method & Perplexity ($\downarrow$) & Accuracy \% ($\uparrow$)\\
		\hline
		Dense & 4.32 & 70.46\\
		\hline
		Window Attention & 5.28 & 67.12\\
		StreamingLLM & 5.30 & 67.02 \\
		H\textsubscript{2}O & 5.08 & 67.69\\
		AutoCompressors & 4.53 & 69.23\\
		${\tool}_P$ & \textbf{4.38}& \textbf{70.35}\\
		\hline
	\end{tabular}
	\label{tab:pretraining}
\end{table}

Note that the training-based methods, such as AutoCompressors and {\tool}, we only update the self-attention-related parameters using LoRA. However, one may argue incorporating additional data could introduce an unfair comparison with other baseline methods. To ensure fairness and further validate the effectiveness of {\tool}, we pre-trained ${\tool}_P$ and other baselines from scratch on the CodeSearchNet dataset and evaluated their performance on language modeling tasks. 

The experimental results are shown in Table~\ref{tab:pretraining}. In this experiment, the KV cache budget is set to 30\%. Notably, under this setting, {\tool} can achieve performance comparable to Dense Attention.
Regarding the training-free, sliding window-based method, it performed well across the evaluation metrics in this experiment, largely due to its design, which prioritizes reducing the KV cache and perplexity in long-text generation, rather than generating contextually coherent code. AutoCompressors demonstrated superior performance in this experiment compared to directly tuning CodeLlama, mainly due to the use of full-parameter fine-tuning in this setting, which enhanced the method's overall effectiveness. This result also supports the efficient fine-tuning capability of {\tool}.
In addition, this experiment can also demonstrate the generalizability of {\tool}, where {\tool} can be applied to models with lower sparsity of attention, as the ${\tool}_P$ is initialized randomly, which prompts the model to allocate attention uniformly across all tokens.


\subsection{RQ2: Memory Consumption of {\tool}}

To address \textbf{RQ2}, we present a comparative analysis of the KV cache consumption (GB) of {\tool} versus that 
of dense attention. Unlike the Sparsity metric, which focuses on the percentage of tokens involvement in computing the attention weights, the evaluation of KV cache overhead necessitates a consideration of the caching strategies employed during the prefilling phase. 

Specifically, dense attention requires the storage of the KV cache for all tokens during the prefilling stage, as these tokens will subsequently serve as context in the decoding phase. 
In contrast, while {\tool} also processes all tokens within the prompt for information aggregation during prefilling, it only needs to cache the KV states for the anchor points. The sparsity metric primarily assesses the number of tokens involved in the computation.

As shown in Table~\ref{tab:cache}, ${\tool}_T$ achieves comparable performance to Dense Attention with an average KV cache requirement of only 0.53 GB. Given the variance in text lengths generated by different methods, we computed the ratio (GB/token) of cache and length to determine the average cache required per token. ${\tool}_T$ maintains performance levels of 102\%, 96\%, and 88\%  when reducing memory overhead by 70\%, 81\%, and 86\%, respectively.

\begin{table}[htbp]
    \centering
    \caption{KV Cache Overhead on HumanEval in RQ2}
    \begin{tabular}{lcccc}
        \hline
        Method & Pass@1 & KV cache (GB)& length & ratio\\
        \hline
        Dense & 31.10 & 1.73  & 55.74 & 3.10$\times$10\textsuperscript{-2}\\
        \hline
        \multirow{3}{*}{${\tool}_T$} & 31.71 & 0.53 ($\downarrow$70\%)& 56.12 & 9.44$\times$10\textsuperscript{-3}\\
        & \cellcolor{gray!30}29.88 & \cellcolor{gray!30}0.33 ($\downarrow$81\%)& \cellcolor{gray!30}56.01 & \cellcolor{gray!30}5.89$\times$10\textsuperscript{-3}\\
        & \cellcolor{gray!50}27.44 & \cellcolor{gray!50}0.24 ($\downarrow$86\%)& \cellcolor{gray!50}54.34 & \cellcolor{gray!50}4.42$\times$10\textsuperscript{-3}\\        
        \hline
    \end{tabular}
	\label{tab:cache}
\end{table}


\subsection{RQ3: Ablation Study} \label{sec:ablation}

For \textbf{RQ3}, we first design ablation experiments that primarily focus on two metrics: Pass@1 and runtime, shown in Table~\ref{tab:ablation}. We then examine the distribution of attention within the model when utilizing LAA to underscore its significance.

\smallskip
\noindent\textbf{Ablation experiments on Pass@1.} The objective of studying the Pass@1 metric is to investigate the impact and contribution of each component within {\tool} evaluating on HumanEval, as shown in Table~\ref{tab:ablation}, the 2nd column. 
\begin{figure}[t]
	\centering
	\begin{minipage}[c]{0.95\linewidth}
		\includegraphics[width=\linewidth]{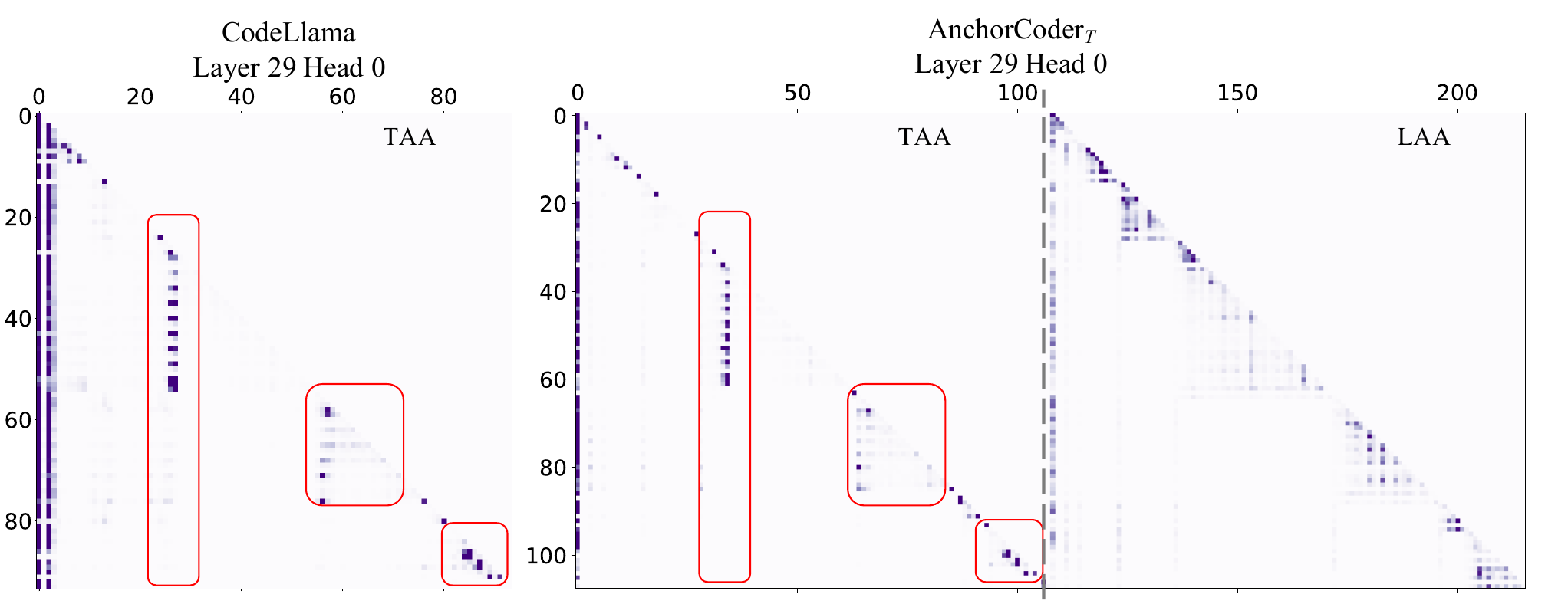}
	\end{minipage}
	\caption{The attention headmap of CodeLlama and ${\tool}_T$.}
	\label{fig:att_compare}
\end{figure}

It can be observed that the removal of LAA leads to a degradation in model performance. This is because the model has limited dimensions, and compression inevitably leads to information loss. LAA serves as an alternative pathway for linking information across layers, significantly reducing this loss of information. In Fig~\ref{fig:att_compare}, we present the attention heatmaps for CodeLlama and ${\tool}_T$, demonstrating that the layers utilizing LAA do not alter the inherent distribution pattern of attention, which ensures the efficiency of tuning. Additionally, a substantial portion of attention is allocated to the KV states in anchor layer, which is crucial for sustaining the overall performance of the model.
MHPE, tailored for anchor points, enhances the model's positional awareness within the overlapping information by adding relative positional information across different attention heads, thereby further augmenting the effectiveness of Anchor Attention.

\smallskip
\noindent\textbf{Ablation experiments on runtime metrics.} We aim to investigate  the temporal cost of the model in generating extended text under different settings. To this end, we configure the generation length to 16,384, which is the training length used by CodeLlama, and present the experimental results in Table~\ref{tab:ablation}, the 3rd-5th column. In this table, we dissect Runtime into three distinct metrics: Prefilling (s), which represents the time the model takes to process the prompt; Decoding (s), which denotes the time taken for autoregressive generation; Throughput (token/s), indicating the number of tokens generated per second during the decoding phase. 

Our analysis reveals that {\tool} can reduce the inference latency of CodeLlama by 25\%, primarily attributed to a bandwidth bottleneck during decoding caused by frequent cache accesses. By reducing the size of the KV cache, we observe a substantial decrease in decoding latency. The incorporation of LAA, serving as a communication channel between layers, increases the computational demands of the model. However, our ablation studies demonstrate that during the decoding phase, latency does not increase significantly, as the presence of a bandwidth bottleneck necessitates the model expending more time accessing the KV cache than calculating attention~\cite{nvidia2022gpu,ivanov2021data,dao2022flashattention}. Consequently, additional computational demands do not substantially contribute to delay. Nevertheless, during the prefilling stage, LAA introduces an additional delay of $3.91 \times 10^{-3}$s (9\%) due to computational bottlenecks at this stage, which is deemed acceptable. Furthermore, our results indicate that MHPE increases inference latency, as it necessitates the computation of distinct positional encodings for each attention head.

\begin{table}[htbp]
    \centering
    \caption{Ablation Result in RQ3}
    \begin{tabular}{lcccc}
        \hline
        \multirow{2}{*}{Model} & \multirow{2}{*}{Pass@1} & \multicolumn{3}{c}{Runtime} \\
        && Prefilling & Decoding & Throughput \\
        \hline
        Dense & $31.10$ & $3.62 \times 10^{-2}$ & $560.39$ & $29.24$ \\
        \hline
        ${\tool}_T$ & $31.71$ & $4.54 \times 10^{-2}$ & $420.35$ & $38.98$ \\

        -w/o LAA & $29.27$ & $4.15 \times 10^{-2}$ & $415.17$ & $39.46$ \\
        -w/o MHPE & $29.88$ & $3.86 \times 10^{-2}$ & $417.28$ & $39.26$ \\
        -TAA only & $28.04$ & $3.71 \times 10^{-2}$ & $411.49$ & $39.82$ \\
    \hline
    \end{tabular}
    \label{tab:ablation}
\end{table}

%% file: sections/7.threat.tex
\section{Threats to Validity} \label{sec:threat}

\noindent\emph{Internal Validity.}  A potential threat to internal validity is the randomness inherent in LLM decoding. To mitigate the impact of uncertainty on experimental results, we employ greedy decoding to ensure the certainty of the generated code by the model, as opposed to the nucleus sampling method commonly used by most LLMs, which introduces randomness.

\smallskip
\noindent\emph{External Validity.} A potential threat to external validity in our study is the generalizability of our findings, which pertains to both language and model types. Our experiments were conducted primarily on Python datasets, which are most commonly required for code generation tasks. Limited by our devices, we conducted experiments only on CodeLlama with a 7B parameters. Notably, we have observed that larger models exhibited sparser attention. Therefore, we believe that larger-scale models have greater potential for compression, which also applies to the method we proposed.

\smallskip
\noindent\emph{Construct validity.} Evaluation metrics pose a potential threat to construct validity for code generation tasks. The metric we selected, pass@k, is the most common and practical in everyday development, compared to the match-based BLEU and CodeBLEU metrics. We also compared sparsity and runtime metrics, which are crucial for assessing the efficiency of code generation. In addition to this, it is important to acknowledge that the accuracy of the pass@k metric in assessing the correctness of generated code is highly dependent on both the quantity and quality of the test cases used. To ensure the comprehensiveness and fairness of testing, we have not only utilized the HumanEval dataset but also employed HumanEvalPlus, which includes over 80 times more test cases compared to HumanEval. This significant increase in the number of test cases allows for a more thorough evaluation of the code's performance in edge cases. This approach ensures that the assessment of generated code is robust and reflects real-world operational challenges more accurately.

%% file: sections/2.background.tex
\section{Related Work} \label{sec:related}

\noindent\emph{Code Generation.}
LLMs have recently demonstrated remarkable capabilities across various applications, particularly in programming-related tasks~\cite{chen2021evaluating, li2022competition}. 
An early standout is Codex~\cite{chen2021evaluating}, which leverages a vast GPT model fine-tuned on GitHub code, fueling the development of Copilot for real-time coding assistance.
Codex has ignited considerable interest in both academia and industry, catalyzing the creation of numerous models. For example, DeepMind's AlphaCode~\cite{li2022competition} is engineered to address coding challenges in competitive programming environments. Similarly, Meta introduced 
models such as InCoder~\cite{fried2022incoder} and CodeLlama~\cite{roziere2023code}, while Salesforce developed CodeRL~\cite{le2022coderl} and CodeGen~\cite{nijkamp2022codegen}. The BigCode project  unveiled StarCoder~\cite{li2023starcoder}. 
In addition, 
numerous open-source large-scale models has further enhanced the capabilities in code generation~\cite{guo2024deepseek,zhu2024deepseek,allal2023santacoder,Wang2023CodeT5OC}.
This surge in model development underscores the significant enhancements in the quality and practicality of automated code generation, marking a substantial leap forward in the methodologies and efficiency with which coding tasks are addressed and accomplished~\cite{wei2023towards,li2020train}.

\smallskip
\noindent\emph{Context Compression.}
The concept of context compression is closely related to 
earlier efforts to archive past representations, enhancing memory and facilitating long-range sequence modeling in Transformers.
Specifically, the Compressive Transformer~\cite{rae2019compressive} utilizes a learned convolutional operator to condense Transformer activations into a more compact memory representation.
Gisting~\cite{mu2024learning} involves training a language model to condense prompts into concise sets of ``gist" tokens, which can be cached and reused to enhance computational efficiency.
AutoCompressors~\cite{chevalier2023adapting} compresses long contexts into compact summary vectors, which are then accessible to the model as soft prompts.
\cite{ren2023context} proposes a plug-and-play approach that can incrementally compress the intermediate activations of a specified span of tokens into more compact forms, thereby reducing both memory and computational costs in processing subsequent contexts.
In contrast to the aforementioned methods, our proposed {\tool} is designed for code generation which compresses in a more natural manner without altering model's inference.

\smallskip
\noindent\emph{Sparse Attention.}
The sparse attention mechanism is a variation of the traditional attention mechanisms used in neural networks, specifically designed to handle large sequences more efficiently by reducing the computational time and memory usage. 
Window attention~\cite{peng2023yarn, chen2023extending} leverages the local features of text for prediction, enabling the model to cache only a minimal amount of KV state for long text predictions. StreamingLLM~\cite{xiao2023efficient} observes the phenomenon of attention sink within large language models and, building upon window attention, introduces sink tokens to enhance long text generation capabilities. H\textsubscript{2}O~\cite{zhang2024h2o} achieves sparse attention by discarding values with low attention scores from the context during decoding, thus maintaining partial model performance with reduced cache requirements. FastGen~\cite{ge2023model} employs a combination of four strategies to effectively restore attention scores and significantly compress the KV cache. Based on these insights, {\tool} utilizing sparse attention in general, reduces the cache overhead via compression rather than extraction, further sustaining model performance.

%% file: sections/8.conclusion.tex
\section{Conclusion} \label{sec:conc}

We have conducted empirical research to explore the sparsity patterns of attention in code generation models. We designed a ``needle in a haystack'' experiment to demonstrate the ineffectiveness of current sparse attention methods in code generation. Based on these findings, we have proposed {\tool}, a novel approach 
which features token-wise and layer-wise anchor attention.
designed to extract and compress the contextual information, and 
mitigate the issues of excessive superposition caused by the compression, respectively. 
Comprehensive experiments have demonstrated that {\tool} significantly reduces the KV cache overhead while maintaining model performance. 

In the future, we plan to extend the application of Anchor Attention to a broader spectrum of attention mechanisms, including but not limited to multi-query attention (MQA)~\cite{shazeer2019fast}, multi-head latent attention (MLA)~\cite{bi2024deepseek}, and grouped-query attention (GQA)~\cite{ainslie2023gqa}. More experiments with larger code LLMs for the repository-level code generation are also planned.   